\documentclass[prper,aps,twocolumn,groupedaddress,floats,showpacs,final,superscriptaddress]{revtex4-1}
\usepackage{leading}
\usepackage[latin3]{inputenc}
\usepackage[makeroom]{cancel}
\usepackage{graphicx}
\usepackage{amsmath}
\usepackage{amsfonts}
\usepackage{amssymb}

\usepackage{color}
\usepackage[left=2cm,right=2cm,top=2cm,bottom=2cm]{geometry}
\usepackage{graphicx} 
\usepackage{dcolumn}
\usepackage{bm}
\usepackage{simplewick}
\usepackage{array}
\usepackage{appendix}

\RequirePackage[
   hyperindex,colorlinks,bookmarksnumbered,
   plainpages=true,pdfstartview=FitH]{hyperref}
\hypersetup{linkcolor=blue,urlcolor=blue,citecolor=blue}
\usepackage{hyperref}
\usepackage{mathrsfs}
\definecolor{purple}{rgb}{0.5,0,0.6}

\usepackage{ulem}
\renewcommand{\emph}[1]{\textit{#1}}%needed, if package  "ulem" is used
\definecolor{darkblue}{rgb}{0,0,0.5}
\definecolor{darkgreen}{rgb}{0,0.5,0}
\definecolor{darkred}{rgb}{.7,0,0}
\definecolor{purple}{rgb}{0.5,0,0.6}
\definecolor{orange}{rgb}{1,0.5,0}
\definecolor{grey}{rgb}{.6,.6,.6}
\definecolor{lightpink}{rgb}{1,0.7,0.75}
\definecolor{pink}{rgb}{1,0.4,0.58}
\definecolor{deeppink}{rgb}{1,0.08,0.58}

\newcommand{\DK}[1]{{\color{black}{#1}}} % Jan 
\renewcommand{\emph}[1]{\textit{#1}}

\newcommand{\mt}{\mathcal{T}}

\newcommand{\ot}{\overline{T}}
\newcommand{\tb}{\overline{T}}
\begin{document}
\title{\DK{Wiedemann-Franz law in scattering theory revisited}}

\author{D. B. Karki}
\affiliation{Division of Quantum State of Matter, Beijing Academy of Quantum Information Sciences, Beijing 100193, China}

\begin{abstract}
The violation of Wiedemann-Franz (WF) law has been widely discussed in quantum transport experiments as an indication of deviation from Fermi-liquid behavior. \DK{The conventional form of WF law is only concerned with the transmission function at Fermi-level which, however, vanishes in many practical situations. We reinvestigate the WF law in noninteracting quantum systems with vanishing zero energy transmission and report a universal number $21/5$ as an upper bound of Lorenz ratio $\mathscr{R}$ in weakly energy-dependent scattering theory. We provide different experimental realizations for the observation of $\mathscr{R}=21/5$ namely the transport setups with graphene, the multi-level quantum dot and double quantum dot. The reported universal Lorenz ratio paves an efficient way of experimentally obtaining the informations about the associated quantum interferences in the system. Our work also provides enough evidence which concludes that the violation of WF law does not necessarily imply the non-Fermi-liquid nature of underlying transport processes; equally, the Fermi-liquid
transport characteristics cannot be concluded by an observed validation of WF law.}
\end{abstract}

\date{\today}
\maketitle

%%%%%%%%%%%%%%%%%%%%%%%%%%%%%%%%%%%%%%% TITLE %%%%%%%%%%%%%%%%%%%%%%%%%%%%%%%%%%%%
Rapid development of quantum technology has stimulated a plethora of quantum transport experiments~\cite{casti}. Thermoelectric phenomena is one of the common transport measurement at the nano scale systems such as in quantum dots, carbon nanotubes and quantum point contacts~\cite{ld1, Zlatic}. 

A prototypical thermoelectric experiment at the nano scale consists of a two terminal device, namely quantum impurity tunnel coupled to two conducting reservoirs. The left (L) and right (R) reservoirs are in equilibrium, separately, at temperatures $T_{\gamma}$ ($\gamma{=}\rm {L, R}$) and chemical potentials $\mu_{\gamma}$ respectively. The heat current ($I_{\rm h}$) and charge current ($I_{\rm c}$) flow across the impurity caused by the temperature gradient $\Delta T\equiv T_{\rm L}{-} T_{\rm R}$ and the mismatch of chemical potentials $\Delta V\equiv\mu_{\rm L}{-} \mu_{\rm R}$. The charge and the heat currents in the linear response theory are connected by the Onsagar relations~\cite{sagar1, sagar2} \DK{which in atomic units read}
\begin{equation}\label{bd0}
 \left(%
\begin{array}{c}
  I_{\rm c} \\
  I_{\rm h} \\
\end{array}%
\right)= \left(%
\begin{array}{cc}
  {\rm L}_{11} & {\rm L}_{12} \\
  {\rm L}_{21} & {\rm L}_{22} \\
\end{array}%
\right)\left(%
\begin{array}{c}
  \Delta V \\
  \Delta T \\
\end{array}%
\right).
\end{equation}

The Onsagar transport coefficients ${\rm L_{\rm ij}}$ in Eq.~\eqref{bd0} provide all the thermoelectric measurements of interests in linear response regime~\cite{costi1}. To this end, we set the transport integrals relating the Onsagar coefficients
\begin{equation}
\mathscr{L}_{\rm n}\equiv \frac{1}{4T}\int^{\infty}_{-\infty}d\varepsilon\;\frac{\varepsilon^n}{\cosh ^2\left(\frac{\varepsilon}{2 T}\right)}\;\mt(\varepsilon, T),\;\;{\rm n}=0, 1, 2.
\end{equation}
Here $T$ is the reference temperature and $\mt(\varepsilon, T)$ is the energy and temperature dependent spectral function (the transmission coefficient).

The transport coefficients characterizing the charge current are expressed in terms of the transport integrals, namely, ${\rm L_{11}}=\mathscr{L}_0$ and ${\rm L_{12}}=-\mathscr{L}_1/T$~\cite{kim}. In addition ${\rm L_{12}}$ and ${\rm L_{21}}$ are related by the Onsagar reciprocity relation and the coefficient ${\rm L_{22}}$ relates the thermal conductance~\cite{casti}. While the electrical conductance is related with $\mathscr{L}_0$ alone the thermopower is usually defined as $\mathcal{S}_{\rm th}=\mathscr{L}_1/\mathscr{L}_0 T$. The thermal conductance $\mathcal{K}$ reads
\begin{equation}
\mathcal{K}=\frac{1}{T}\left[\mathscr{L}_{\rm 2}-\frac{\mathscr{L}_{\rm 1}^2}{\mathscr{L}_{\rm 0}}\right].
\end{equation}
In addition, the Wiedemann-Franz (WF) law connects the electronic thermal conductance $\mathcal{K}$ to the electrical conductance $G$ in low temperature regime of a macroscopic sample by an universal constant, the Lorenz number $L_0$, defined as $L_0\equiv \mathscr{K}/GT=\pi^2/3$. The constant value of Lorenz number simply implies that the transport mechanisms responsible for heat and charge currents are fundamentally the same~\cite{costi1}.

The possible deviation from WF law at the nano scale has been accounted for by considering the Lorenz ratio~\cite{jukka}.
\begin{equation}
\mathscr{R}\equiv\frac{L(T)}{L_0}=\frac{3}{(\pi T)^2}\left[\frac{\mathscr{L}_{\rm 2}}{\mathscr{L}_{\rm 0}}-\left(\frac{\mathscr{L}_{\rm 1}}{\mathscr{L}_{\rm 0}}\right)^2\right],
\end{equation}
the deviation of $\mathscr{R}$ from unity amounts the violation of WF law. Although the WF law is expected to be violated strongly at the nano scale, surprisingly it \DK{works} quantitatively well for $T\to 0$ even for some interacting systems \DK{with both Fermi-liquid and non Fermi-liquid correlations} such as the Kondo correlated systems~\cite{costi1, mkk, dbk}. \DK{This suggests that the Fermi-liquid nature of transport can not be concluded by the observed validation of WF law.  In addition, it might be also possible that the quantum transport in Fermi-liquid regime (for both interacting and noninteracting systems) strongly violates the WF law. To explore this possibility in detail, we restrict ourselves by considering the noninteracting systems described by the scattering theory. To this end}, we sketch briefly the main assumption behind the derivation of \DK{original} WF law and provide the logical reason of relaxing such assumption \DK{which eventually results in the different Lorenz ratio from the conventional one.} We consider the transmission function satisfying the condition~\cite{whitney1, whitney2}
\begin{equation}
0\leq\mt(\varepsilon, T)\leq N,
\end{equation}
with $N$ being the number of transverse conduction modes. In addition, for the system modelled by the scattering theory the transmission function is merely energy dependent, the temperature comes solely from the Fermi-function $\mt(\varepsilon, T)=\mt(\varepsilon)$~\cite{whitney1}. The fundamental assumption of obtaining WF law at nano scale is to consider the smooth transmission function such that
\DK{\begin{equation}\label{aama1}
\mt(\varepsilon)\simeq\mt_0+\left.\frac{\partial\mt(\varepsilon)}{\partial\varepsilon}\right|_{\varepsilon=0}\;\varepsilon+\cdots,
\end{equation}}
where $\mt_0$ is the zero energy transmission function. Assuming the unitary condition $\mt(\varepsilon=0)\to 1$, one can readily sees that $\mathscr{L}_0=1$, $\mathscr{L}_2=(\pi T)^2/3$ and $\mathscr{L}_1\simeq 0$ satisfying the condition provided by WF law $L_0=\pi^2/3$.

Toward the urge for enhancing thermoelectric performance, the concept of ideal energy
filters with an energy-dependent transmission function $\mt(\varepsilon)\propto \Gamma\delta(\varepsilon-\varepsilon_0)$ has been suggested, with $\varepsilon_0$ being the position of single level contributing the transport and $\Gamma$ is some energy scale of the system~\cite{hh2}. This delta function transmission has then realized being of not much practical application since this abruptly reduces the efficiency at maximum power~\cite{hh3}. As an overcome of this difficulty, two-level systems have been proposed where the quantum interference can substantially improve
the maximum thermoelectric power and the efficiency at maximum power~\cite{hh}. We note that the quantum interference between different conduction channels is quite common phenomena at the nano scale including the strongly correlated systems~\cite{GP_Review_2005, dee2, dee3}. In the presence of two or more interfering transport channel, the fundamental assumption of WF law that $\mt(\varepsilon=0)\to 1$ is completely violated, rather the destructive interference results in $\mt(\varepsilon=0)\to 0$. In addition, the system might posses particle-hole (PH) symmetry on the top of level interference, that is $\mt(\varepsilon)=\mt(-\varepsilon)$. \DK{These two properties can also be observed in graphene since the associated unusual band structure~\cite{jim,HX00, HX0, HX}}.

For these two different practical \DK{cases presented above, the transport through the multi-level quantum dot (QD) or multi-QD and the graphene, the crude assumption of smooth transmission function given by Eq.~\eqref{aama1} is not of practical use}. In this case one has to rather go beyond the first order expansion of transmission coefficient (see the following section). The fundamental question of paramount importance, both theoretical and experimental interest, then would be what about the WF law for the systems possessing $\mt(\varepsilon=0)\to 0$ and $\mt(\varepsilon)=\mt(-\varepsilon)$. More generally, what is the connection between the electronic thermal conductance and the electrical conductance for the systems with graphene like transmission?

\DK{To unveil the form of WF law with vanishing zero energy transmission}, we consider a generic noninteracting system which is well described within the scattering theory. \DK{Since many experimental systems posses rather weak energy dependence of their transmission~\cite{casti}, we express the transmission function into the Taylor series in energy}
\begin{equation}\label{aama2}
\mt(\varepsilon)=\mt_0+\mt_1\;\frac{\varepsilon}{\Gamma}+\mt_2\left(\frac{\varepsilon}{\Gamma}\right)^2+\cdots,
\end{equation}
where $\mt_0$ is the zero energy transmission and $\mt_{1, 2}$ are the expansion coefficients. \DK{The truncation of the series Eq.~\eqref{aama2} at the second order is indeed the reasonable approximation for most of the practical situations unless the exotic situation with $\mt_{0, 1, 2}=0$ is encountered. Although, the simultaneous vanishing of $\mt_0$, $\mt_1$ and $\mt_2$ is very unlikely to be the case of real experiment, we will revisit this case later and for now we truncate the series Eq.~\eqref{aama2} at second order. The Lorenz ratio obtained from the Eq.~\eqref{aama2} then reads}
\begin{equation}\label{aama3}
\mathscr{R}=\frac{3}{5} \left[7-\frac{16 \mt_0 \left(\pi ^2 \tb^2 \mt_2+3 \mt_0\right)+5 \pi ^2 \tb^2 \mt_1^2}{\left(\pi ^2 \tb^2 \mt_2+3 \mt_0\right)^2}\right]+\cdots
\end{equation}
with $\ot\equiv T/\Gamma$. In addition we have used the Sommerfeld integrals $\mathcal{I}_j\equiv\frac{1}{4T}\int^{\infty}_{-\infty}d\varepsilon\;\frac{\varepsilon^j}{\cosh^2\left(\varepsilon/2T\right)}$ with $\mathcal{I}_0=1,\;\mathcal{I}_2=(\pi T)^2/3$ and $\mathcal{I}_4=7/15{\times}(\pi T)^4$ and vanishing odd integrals. For single level transport where $\mt_0\neq 0$, one can expand Eq.~\eqref{aama3} in low temperature limit to get
\begin{equation}
\left.\mathscr{R}\right|_{\mt_0\neq 0}=1+(\pi \ot)^2\left[\frac{16}{15}\frac{\mt_2}{\mt_0}-\frac{1}{3}\left(\frac{\mt_1}{\mt_0}\right)^2\right]+\cdots,
\end{equation}
which immediately verifies the WF for $T\to 0$. For the systems possessing the special property $\mt_0=\mt(\varepsilon\to 0)=0$ we have, however, quite different form of Lorenz ratio~\footnote{From the symmetry property of the transport integrals, the vanishing $\mt_0$ also implies the vanishing $\mt_1$. Therefore, the contribution to the thermopower comes from the terms containing $\mt_3$ and beyond in Eq.~\eqref{aama2}. This fact is apparent from the general treatment of the quantum interferences resulting in the Eq.~\eqref{aama10} and also from the explicit calculations involving the multilevel QD and double-QDs.}
\begin{equation}\label{aama4}
\left.\mathscr{R}\right|_{\mt_0=0}=21/5-\mathcal{S}^2_{\rm th}/L_0+\cdots
\end{equation}
\DK{In scattering theory (or systems described by the Fermi-liquid paradigm) thermopower posses a linear temperature scaling behavior which gets vanishes for $T\to 0$. Therefore the zero temperature limit of Lorenz ratio reads
\begin{equation}\label{baaaama}
\left.\mathscr{R}\right|_{\mt_{0}= 0,\; T\to 0}=\frac{21}{5},\;\;\DK{L=\frac{7\pi^2}{5}}.
\end{equation}}
This results suggests that the presence of quantum interference or a system with rather special geometry with vanishing zero energy transmission strongly violates the WF law. In this limit the Lorenz ratio attains the temperature independent universal number $21/5$.

\DK{The prediction of universal number $21/5$ for the Lorenz ratio originated mainly form the assumption of vanishing zero energy transmission $\mt_0=0$. The contribution of $\mt_1$ to Lorenz ratio eventually vanishes even if the PH symmetry is explicitly broken by taking the limit of $T\to 0$ (thermopower possesses a linear scaling with temperature). Therefore the consideration of only non-vanishing element $\mt_2$ is the sole reason of observed universal Lorenz ratio. Although it is very unlikely that a real system possesses the property of $\mt_{0, 1, 2}\to 0$, it is also worth of commenting on this rather exotic situation. In this case the Eq.~\eqref{aama2} rather starts from the third order term in the energy expansion and the Lorenz ratio attains rather a big number of $465/49$ at the limit of $T\to 0$. With these observations, one can conclude that there is no unique upper bound for the Lorenz ratio in scattering theory. Nevertheless, most of the experimental systems have reasonably weak energy dependence of their transmission and hence the predictions made in our work based on Eq.~\eqref{aama2} have rather broad domain of applicability.}
\begin{figure}
\includegraphics[scale=0.2]{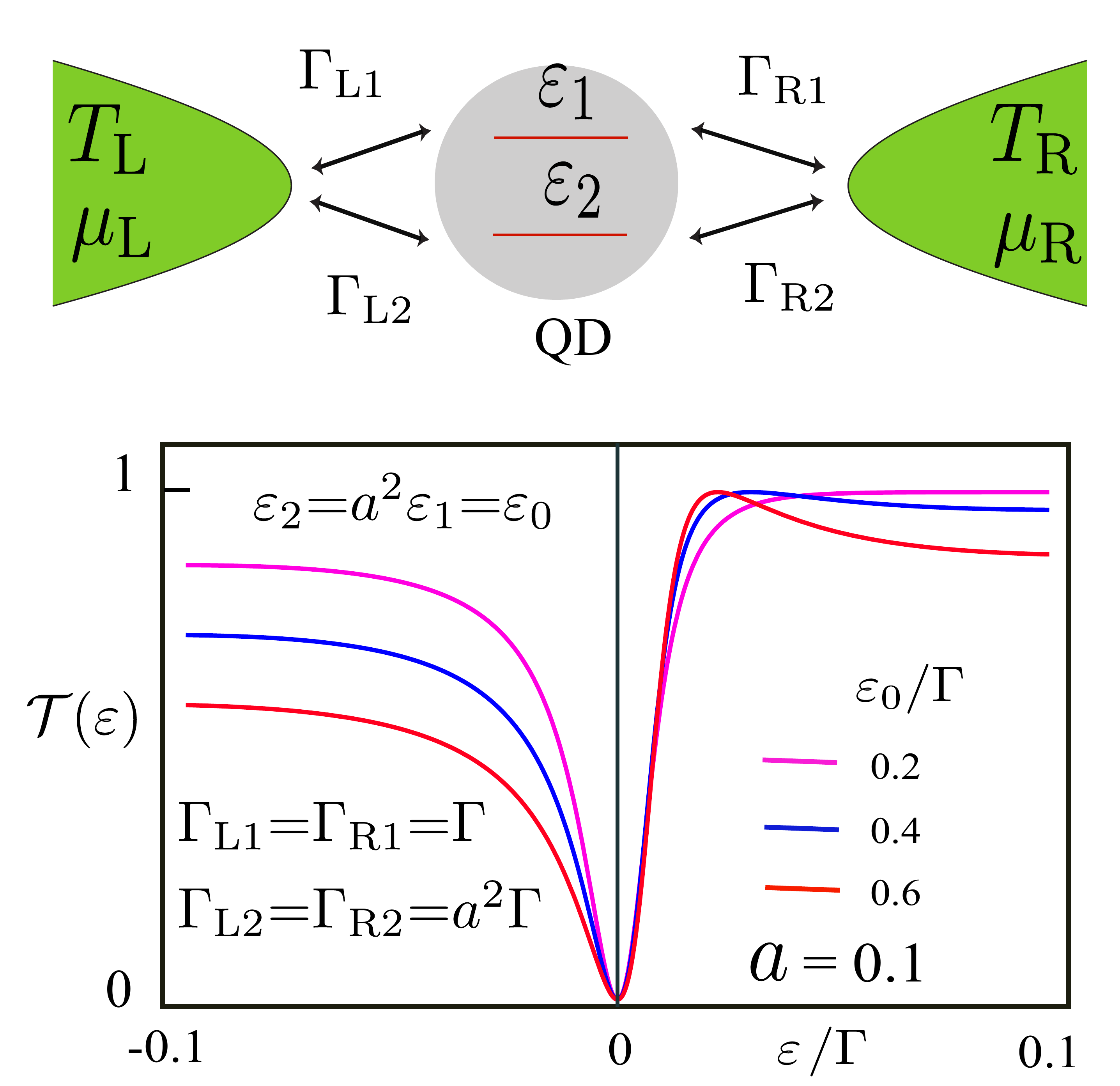}
\caption{\DK{Upper panel: The multi-level QD transport setup. A QD with two-levels $\varepsilon_i$ ($i{=}1, 2$) is tunned coupled to the left (L) and right (R) electronic reservoirs. The symbol $\Gamma_{\alpha i}$ ($\alpha{=}{\rm L, R}$) stands for the coupling strength between the lead $\alpha$ and the energy level $\varepsilon_i$ in the QD. We assume that the tunneling does not mix the two levels which is valid in the noninteracting transport setups. Lower panel: The vanishing zero energy transmission function in multi-level QD with properly chosen coupling strengths $\Gamma_{\alpha i}$ and level positions $\varepsilon_i$. }\label{deep1}}
\end{figure}

\begin{figure}[h]
\begin{center}
\includegraphics[scale=0.8]{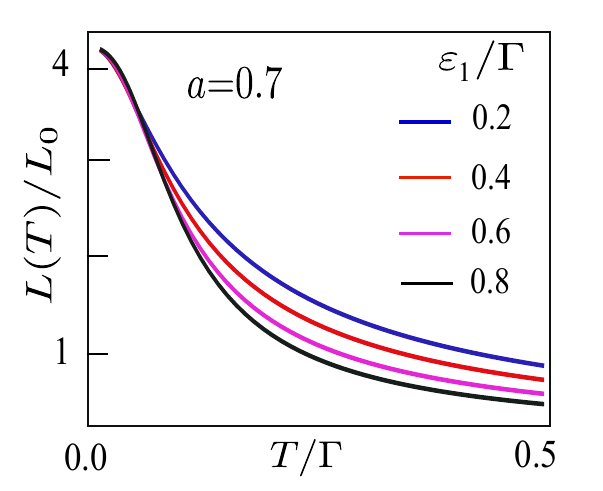}
\caption{\DK{The variation of Lorenz ratio $\mathscr{R}=L(T)/L_0$ with temperature $T$ for given positions of energy level $\varepsilon_1$ in two-level quantum transport setup at fixed parity parameter $a$ (see Eq.~\eqref{aggr1} and following text for the details).}}\label{aama}
\end{center}
\end{figure}

The graphene based transport would be one of the simplest example to \DK{experimentally verify} the predicted universal \DK{number for Lorenz ratio}~\cite{HX}. In the following we further corroborate on the \DK{ubiquitousness of predicted Lorenz ratio in quantum experiments} and study the more general behavior of Eq.~\eqref{aama4}. To this end, we consider a two-level quantum dot tunnel coupled to two conducting leads (reservoirs) \DK{as shown in Fig.~\ref{deep1}}. In addition, for the activation of quantum interference we assume that the two levels $\varepsilon_{1, 2}$ couple with different parity to
the leads and their coupling strengths differ by a factor $a^2$, $\Gamma_{\rm L1}=\Gamma_{\rm R1}=\Gamma$, $\Gamma_{\rm L2}=\Gamma_{\rm R2}=a^2\Gamma$ (see Ref.~\cite{hh3} for detail). In this case the transmission function $\mt_{\rm M}(\varepsilon,\varepsilon_1,\varepsilon_2)$ reads
\begin{equation}
\mt_{\rm M}(\varepsilon,\varepsilon_1,\varepsilon_2)=\Gamma^2\Big|\frac{1}{\varepsilon-\varepsilon_1+i\Gamma}-\frac{a^2}{\varepsilon-\varepsilon_2+ia^2\Gamma}\Big|^2.
\end{equation}
The zero energy transport will be nullified for a particular choice of $\varepsilon_2=a^2\varepsilon_1$, that is $\mt_{\rm M}(\varepsilon\to 0,\varepsilon_1,\varepsilon_2\to a^2\varepsilon_1)\to 0$. For this case of $\varepsilon_2=a^2\varepsilon_1$ we recast the transmission function into the form\DK{
\begin{equation}\label{aggr1}
\mt_{\rm M}(\tilde{\varepsilon}, \tilde{\varepsilon}_0){=}\frac{\left(a^2{-}1\right)^2 \tilde{\varepsilon}^2}{\left(1{+}(\tilde{\varepsilon}{-}\tilde{\varepsilon}_0)^2\right) \left(a^4 \left(1{+}\tilde{\varepsilon}_0^2\right){-}2 a^2 \tilde{\varepsilon} \tilde{\varepsilon}_0{+}\tilde{\varepsilon}^2\right)},
\end{equation}
where energies are expressed in the unit of $\Gamma$, that is $\varepsilon/\Gamma\equiv\tilde{\varepsilon}$ and $\varepsilon_1/\Gamma=\varepsilon_0/\Gamma\equiv\tilde{\varepsilon}_0$.} For this transmission function, in the limit of $T\to 0$ we recover again the universal Lorenz ratio of $21/5$ irrespective of the parameters $\tilde{\varepsilon}_0$ and $a$. This can be easily seen by considering the above transmission shape at the strong coupling regime, $(\tilde{\varepsilon}, \tilde{\varepsilon}_0)\ll 1$, which within the lowest order expansion reads
\begin{equation}
\mt_{\rm M}(\tilde{\varepsilon})=\left(a^2-1\right)^2\tilde{\varepsilon}^2/a^4+\cdots
\end{equation}
The violation of WF law with varying temperature for the two level system consider above is as shown in Fig.~\ref{aama}, \DK{ which apparently reaches the universal number of $21/5$ for the Lorenz ratio at the limit of $T\to 0$}. 

\begin{figure}
\includegraphics[scale=0.2]{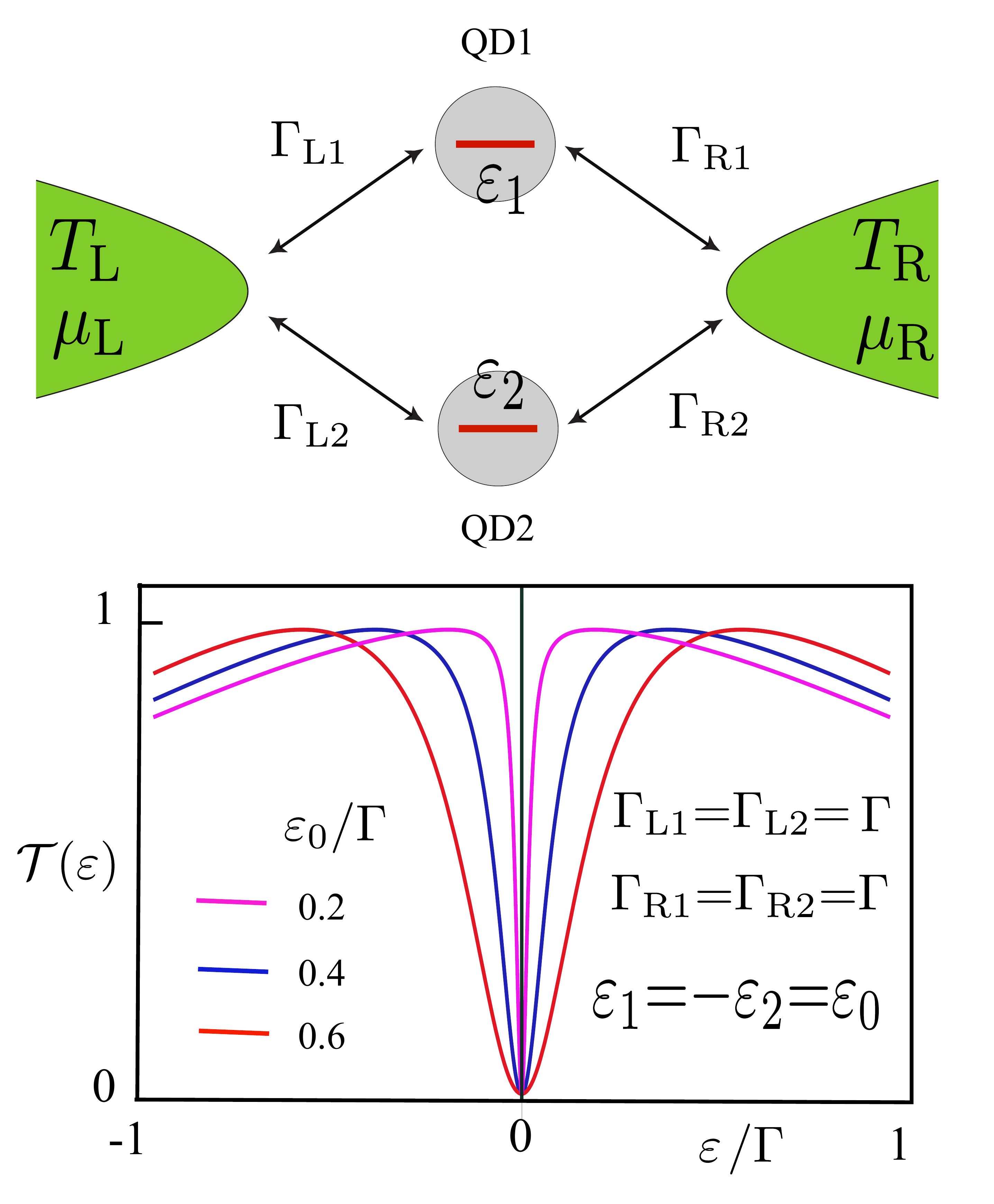}
\caption{\DK{Upper panel: The DQD transport setup setup. Two QDs (QD1 and QD2) each with single energy level $\varepsilon_i$ ($i=1, 2$) are tunnel coupled to the left and right electronic reserviors in the parallel configuration. $\Gamma_{\alpha i}$ represents the coupling strength between the lead $\alpha$ and the $i^{\rm th}$ QD. Lower panel: The energy dependence of the transmission function for the DQD setup in the upper panel which exhibits the anti-resonance provided the specific choice of parameters satisfying $\Gamma_{\alpha i}=\Gamma$ and $\varepsilon_1=-\varepsilon_2=\varepsilon_0$.}}\label{deep2}
\end{figure}

For further discussion on the \DK{experimental verification of Lorenz ratio}, we consider a double quantum dot setup with equal tunneling amplitudes $\Gamma$ and respective energy levels $\varepsilon_{1, 2}$ \DK{as shown in Fig.~\ref{deep2}}. Tuning the system in the regime of $\varepsilon_1=-\varepsilon_2=\varepsilon_0$, the double quantum dot (DQD) transmission function $\mt_{\rm DQD}(\varepsilon)$ reads~\cite{kbd4, kbd3, kbd2, kbd1, kom}\DK{
\begin{equation}\nonumber
\mt_{\rm DQD}(\tilde{\varepsilon}){=}\frac{1}{\sqrt{1{-}\tilde{\varepsilon}_0^2}}\left[\!\frac{\Omega_+^2}{\tilde{\varepsilon}^2{+}\Omega_+^2}{-}\frac{\Omega_-^2}{\tilde{\varepsilon}^2{+}\Omega_-^2}\right]\!,\Omega_{\pm}{=}1{\pm} \sqrt{1{-}\tilde{\varepsilon}_0^2},
\end{equation}}
where we expressed the energy in the unit of $\Gamma$ satisfying the condition $1>\varepsilon_0/\Gamma\equiv\tilde{\varepsilon}_0$. Since the low energy expansion of the function $\mt_{\rm DQD}(\tilde{\varepsilon})$ has the simple form $\mt_{\rm DQD}(\tilde{\varepsilon})=4 \left(\tilde{\varepsilon}/\tilde{\varepsilon}_0\right)^2+\cdots$,  the universal number of $21/5$ for the Lorenz ratio is exactly recovered at $T\to 0$.

\DK{We note that one of the fundamental factor validating above discussions is also the quantum interference effects. Testing our predictions with a multi-level QD, therefore, only needs the proper tuning of parity factor $a$ as seen from the Eq.~\eqref{aggr1}. Besides, for the study of quantum interferences in DQD setup, the relative phases of the tunneling amplitudes may represent an Aharonov-Bohm flux~\cite{refadd1}. The detail study of the WF law in DQD setup with Aharonov-Bohm (AB) geometry is left for the future work. It can be also the case that dot structure possesses the small area so that it does not really generate the AB-phase~\cite{kom}. Given that, however, the destructive interference (antiresonance) in DQD setup resulting in $\mt_0\to0$ needed to verify our predictions can be achieved by the appropriate
choice of the gate voltage. }

For the propose of strengthening our prediction, in the following we formulate the multi-level transport description using more general scattering matrix formulation. In two terminal transport description, the electron operators in the left and right leads $c_{\rm L, R}$ can always be rotated using the Glazman-Raikh rotation~\cite{GR} (we consider the symmetric lead-dot coupling)
\begin{equation}\label{sane5}
 \left(%
\begin{array}{c}
  b_e\\
  b_o\\
\end{array}%
\right)= \mathbb{U}\left(%
\begin{array}{c}
  c_{{\rm L}} \\
  c_{{\rm R}} \\
\end{array}%
\right),\;\;\mathbb{U}\equiv\frac{1}{\sqrt{2}}\left(%
\begin{array}{cc}
 1 & \phantom{-}1 \\
  1 & -1 \\
\end{array}%
\right).
\end{equation}
For the case of conventional single level transport, the transformation Eq.~\eqref{sane5} effectively decouples the operators $b_0$ from the impurity degrees of freedom~\cite{dee1, last1}. Multi-level setups, however, result in the coupling of both operators $b_{e, o}$ with quantum dot degrees of freedom~\cite{dee2}. For simplicity we consider just two levels both of them are close to the resonance scattering. The resonance phenomena are generally described in terms of scattering phase shifts $\delta_{e,o}$ occurring in interfering channels. Since both channels $b_o$ and $b_e$ are close to the resonance values, the phase shift occurring in the respective channels is accounted for by the scattering matrix $\mathbb{S}_{\rm tot}$
\begin{equation}
\mathbb{S}_{\rm diag}=\left(%
\begin{array}{cc}
e^{2i\delta_e} & 0 \\
  0 & e^{2i\delta_o} \\
\end{array}%
\right),\;\;\mathbb{S}_{\rm tot}=\mathbb{U}^{\dagger}\mathbb{S}_{\rm diag}\mathbb{U}.
\end{equation}
For the purely scattering effects, the transmission function can then be obtained as~\cite{dee2}
\begin{equation}\label{yue}
\mt(\varepsilon)=\Big|\mathbb{S}_{\rm tot}\Big|^2=\sin^2\left(\delta_e-\delta_o\right).
\end{equation}
Energy dependence of phase shifts $\delta_{e, o}$ come from the expansion with some arbitrary constants $\alpha_{e, o}$ and $\beta_{e, o}$
\begin{equation}\label{mmjk}
\delta_{e, o}(\varepsilon)=\delta_0+\alpha_{e, o}\left(\varepsilon/\Gamma\right)+\beta_{e, o}\left(\varepsilon/\Gamma\right)^2+\cdots,
\end{equation}
where $\delta_0$ is the zero energy phase shift which can be set at same level for all the interfering channels~\footnote{If $\delta_0$ is chosen to be different for the even channel and odd channel, $\delta_{0, e}\equiv \delta^P_e$ and $\delta_{0, o}\equiv \delta^P_o$ with $\delta^P_e\neq \delta^P_o$, the transmission function acquires the finite $\mt_0$ term: $
\mt(\varepsilon)=\sin^2\left(\delta^P_{e}-\delta^P_{o}\right)+
\left(\alpha_e-\alpha_o\right)\sin2\left(\delta^P_{e}-\delta^P_{o}\right)\varepsilon+
\left(\alpha_e-\alpha_o\right)^2\cos2\left(\delta^P_{e}-\delta^P_{o}\right)\varepsilon^2
+\mathcal{O}(\varepsilon^3)$. In this case the WF law is exactly satisfied at $T\to 0$.}. Therefore, the low energy form of transmission coefficient reads
\begin{equation}\label{aama10}
\mt(\varepsilon)=\left(\alpha_e-\alpha_o\right)^2\;\left(\varepsilon/\Gamma\right)^2.
\end{equation}
The transmission coefficient expressed in Eq.~\eqref{aama10} is very generic which always \DK{attains the universal value $21/5$ of the Lorenz ratio. We note that the limiting case of $\alpha_e\to\alpha_o$ represents the complete destructive interferences between the channels participating in the transport process. At this limiting case the transport is completely blocked. In the absence of channel mixing (noninteracting systems), the Eq.~\eqref{yue} is an exact result showing, in an unified way, the importance of quantum interferences in generic transport experiments.} 

\DK{In conclusion, we reported an upper bound of the Lorenz ratio in the systems possessing weak energy-dependent transmission. In particular, we investigated the quantum transport through the noninteracting systems with vanishing zero energy transmission where the main hypothesis of conventional WF law must be relaxed. The vanishing of zero energy transmission has been explored for many systems of experimental interests and found to be associated either with the quantum interference effects or the internal structure of the system. In this case of vanishing zero energy transmission, the Lorenz ratio attains an universal number of $\mathscr{R}=21/5$ at $T\to 0$ which is significantly higher than that predicted withing conventional WF law. This universal number, therefore, provides an experimental way of characterizing the interference phenomena in noninteracting nano scale devices. We observed that there exists no finite upper bound for the Lorenz ratio in scattering theory. Nevertheless, for many experimental systems the Lorenz ratio predicted in this work might even serve as an upper bound. Our prediction can easily be verified in graphene in the absence of screening effects and considering only the long ranged impurity. In addition, we propose the noninteracting multi-level quantum dot and double quantum dots setups as the ideal candidates to test our predictions. Our
work provides significant evidence which concludes that the
violations of WF law do not necessarily imply a non-Fermi-
liquid nature of the underlying transport phenomena; equally,
Fermi-liquid transport characteristics cannot be concluded by
an observed validation of WF law.}

We are thankful to Hong-Yi Xie for fruitful discussions and drawing our attention to the Ref~\cite{HX}. We are also grateful to Mikhail Kiselev for illuminating discussions.
%\bibliography{WF_law}

\end{document}